\documentstyle[aps,preprint]{revtex}
\begin{document}
\draft
\title{Coupled eigenmodes in a two-component 
Bose-Einstein condensate}
\author{Patrik \"Ohberg and Stig Stenholm}
\address{Department of Physics, Royal Institute of Technology}
\address{Lindstedtsv\"agen 24, S-10044 Stockholm} 
\address{SWEDEN}

\date{\today}
\maketitle

\begin{abstract}
We have studied the elementary excitations in a two component Bose-Einstein 
condensate. We concentrate on the breathing modes and find the elementary 
excitations to possess avoided crossings and regions of coalescing 
oscillations where both components of the condensates oscillate with the 
same frequency. For large repulsive interactions between the
condensates, their oscillational modes tend to decouple due to
decreased overlap. A thorough investigation of the eigenmodes near
the avoided crossings are presented. 
\end{abstract}

\pacs{03.75.Fi,05.30.Jp}

\section{Introduction}

Bose-Einstein condensation in trapped dilute gases differs from the bulk case
 in the lack of translational invariance \cite{Cor95,Ket95,Hul95}. At zero 
temperature the ensuing spatially varying condensates can be calculated 
reliably from the Gross-Pitaevskii equations \cite{KBG4}. At finite 
temperatures the region close to the transition point has been treated by a 
Hartree-Fock scheme \cite{Griff,me3}.

In order to obtain the spectrum of the condensed gases a variety of methods 
have been applied, most of them utilizing a Bogoliubov-de Gennes 
diagonalization of the dynamics linearized around the ground state 
\cite{Str,me1,KBG,Sing,You}, and the eigenmodes and their corresponding 
frequencies are 
well known \cite{Jin96,Mew96,Mew97,Jin97,Kurn98}. 
The calculations of the temperature 
dependence of the excitations have still not reached decisive conclusions 
\cite{Sam,Pita,Liu,Fedi,Giorgi}.

A new physical situation emerged when it was found that two hyperfine 
components, $|F=2,m=2>$ and $|F=1,m=-1>$, of Rb can be induced to condense
concomittantly in a single trap \cite{Myat}. The two condensates see different 
potentials, and they tend to appear displaced from each other thanks to the 
effect of gravity. This is however, a technical feature only and it is
possible to make the condensates in a totally symmetric environment
\cite{symm} . In 
such a situation, we have shown \cite{me2} by a Hartree-Fock treatment that 
the symmetrically formed condensate may find a state of lower energy by 
breaking the symmetry. If the experimentally realized condensate falls into 
the parameter range where this ooccurs remains to be seen.

In this paper we discuss the oscillational spectrum of a symmetric double 
condensate. Other works investigating the low lying excitations of the double 
condensate use a variational approach \cite{Busch} and a direct numerical 
treatment \cite{Esry,Big2,Gord}. We linearize
the Gross-Pitaevskii equations of the double condensate and solve for the 
eigenmodes. Because of the numerical burden of the computations, we only 
look at the radially symmetric $(l=0)$ oscillations, the breathing modes 
\cite{Jin96,Mew97}. We have used time dependent integration methods to check 
the results of our eigenmode analysis, the details will be explained below. As 
results we have found the oscillation modes to display avoided crossings as 
expected of quantum energy levels.

The organization of the present paper is as follows. Section II
investigates  the small oscillations of the coupled Gross-Pitaevskii 
equations for the double 
condensate. Their dynamics is linearized and the usual Bogoliubov-de Gennes 
expansion \cite{KBG2,Fetter} is generalized to the two-component condensate 
case, which gives an eigenvalue problem, which is nonhermitian, but has real 
eigenvalues because of the time reversal symmetry of the problem. In Sec.III
 we explain how to solve this numerically, and present methods to verify the
 results  by direct time integration of the equations when perturbed from 
equilibrium. The results of the calculations are presented and discussed  in 
Sec. IV. The main conlusions are presented in Sec. V.

\section{Formulating the problem}

We consider a two-component Bose condensed gas in external harmonic
potentials 
\begin{equation}
V_{i}(r)={1\over 2} m \Omega_{i}^{2} r^{2} \qquad i=1,2
\end{equation}
with the trap frequencies $\Omega_{i}$ and assume a pair potential of the
atom-atom interaction of the form $V_i =v_i \delta ({\bf r-r'})$ with 
\begin{equation}
v_{i}={{4 \pi \hbar^{2} a_{i}}\over {m}} \qquad i=1,2,3
\end{equation}
where $a_{1}$ and $a_{2}$ stand for the inter species scattering lengths and 
$a_{3}$ is the scattering length between the two different atoms, and
$m$ is the mass of the atoms which is taken to be the same for both
species. The grand canonical Hamiltonian can then be written in the form
\begin{eqnarray}
\hat H &=& \int d{\bf r}\Bigl\{ \hat \Psi_{1}^{\dagger}({\bf r}) 
[ -{{\hbar^{2}}\over
{2m}} \nabla^{2}+V_{1}(r)-\mu_{1}]\hat \Psi_{1}({\bf r})+
\hat \Psi_{2}^{\dagger}({\bf r}) [ -{{\hbar^{2}}\over
{2m}} \nabla^{2}+V_{2}(r)-\mu_{2}]\hat \Psi_{2}({\bf r})\nonumber \\ &&
+ {1\over 2} v_{1} \hat \Psi_{1}^{\dagger}({\bf r}) \hat
\Psi_{1}^{\dagger}({\bf r}) \hat \Psi_{1}({\bf r}) \hat
\Psi_{1}({\bf r}) +
{1\over 2} v_{2} \hat \Psi_{2}^{\dagger}({\bf r}) \hat
\Psi_{2}^{\dagger}({\bf r}) \hat \Psi_{2}({\bf r}) \hat          
\Psi_{2}({\bf r})  \nonumber \\ &&
+v_{3} \hat \Psi_{1}^{\dagger}({\bf r}) \hat \Psi_{2}^{\dagger}({\bf r})
\hat \Psi_{1}({\bf r}) \hat \Psi_{2}({\bf r}) \Bigr\} \label{ham1}
\end{eqnarray}
where $\hat\Psi_{i}({\bf r})$ is the field operator of the bosons and 
$\mu_{i}$ 
is the chemical potential. To include this in the Hamiltonian only shifts the 
energies, and introduces the dependence on the particle number directly into 
the Hamiltonian.

Using the canonical bosonic commutation relations between the fields $\{ \hat
\Psi_i \}$ we obtain the equations
\begin{eqnarray}
i \hbar {\partial \over {\partial t}}\hat\Psi_{1}({\bf r})&=&\bigl 
[ -{{\hbar^{2}}\over
{2m}} \nabla^{2}+V_{1}(r)+v_{1}\hat\Psi_{1}^{\dagger}({\bf r})
\hat\Psi_{1}({\bf r})+v_{3} \hat\Psi_{2}^{\dagger}({\bf r})
\hat\Psi_{2}({\bf r})-\mu_1 \bigr ]\hat\Psi_{1}({\bf r}) 
\label{tggp1} \\
i \hbar {\partial \over {\partial t}}\hat\Psi_{2}({\bf r})&=&\bigl [ 
-{{\hbar^{2}}\over
{2m}} \nabla^{2}+V_{2}(r)+v_{2}\hat\Psi_{2}^{\dagger}({\bf r})
\hat\Psi_{2}({\bf r})+v_{3}\hat\Psi_{1}^{\dagger}({\bf r})
\hat\Psi_{1}({\bf r})-\mu_2 \bigr ]\hat\Psi_{2}({\bf r}). 
\label{tggp2}
\end{eqnarray}

Taking the expectation values and assuming the products to factorize, we obtain
the time dependent two-component Gross-Pitaevskii equations
\begin{mathletters}
\label{tgp}
\begin{eqnarray}
i \hbar {\partial \over {\partial t}}\Psi_{1}({\bf r})&=&\bigl 
[ -{{\hbar^{2}}\over
{2m}} \nabla^{2}+V_{1}(r)+v_{1}|\Psi_{1}({\bf r})|^{2}+v_{3}
|\Psi_{2}({\bf r})|^{2}-\mu_1 \bigr ]\Psi_{1}({\bf r}) 
\label{tggp3} \\
i \hbar {\partial \over {\partial t}}\Psi_{2}({\bf r})&=&\bigl [ 
-{{\hbar^{2}}\over
{2m}} \nabla^{2}+V_{2}(r)+v_{2}|\Psi_{2}({\bf r})|^{2}+v_{3}
|\Psi_{1}({\bf r})|^{2} -\mu_2 \bigr ]\Psi_{2}({\bf r}). 
\label{tggp4}
\end{eqnarray}
\end{mathletters}
where the chemical potentials are chosen such that 
\begin{equation}
\int d{\bf r}|\Psi_{i}({\bf r})|^{2}=N_{i}\qquad i=1,2.
\end{equation} 

When we set the time derivatives in Eqs. (\ref{tggp3}) and (\ref{tggp4}) 
equal to zero, we obtain the Gross-Pitaevskii ground state functions 
$\{ \Psi_1^g,\Psi_2^g\}$ which we assume to be chosen real. They solve the 
equations 
\begin{mathletters}
\label{ggp}
\begin{eqnarray}
\mu_{1}\Psi_{1}^{g}({\bf r})&=&\bigl [ -{{\hbar^{2}}\over
{2m}} \nabla^{2}+V_{1}(r)+v_{1}\Psi_{1}^{g}({\bf r})^{2}+v_{3}
\Psi_{2}^{g}({\bf r})^{2}\bigr ]\Psi_{1}^{g}({\bf r}) \label{ggp1} \\
\mu_{2}\Psi_{2}^{g}({\bf r})&=&\bigl [ -{{\hbar^{2}}\over
{2m}} \nabla^{2}+V_{2}(r)+v_{2}\Psi_{2}^{g}({\bf r})^{2}+v_{3}
\Psi_{1}^{g}({\bf r})^{2}\bigr ]\Psi_{2}^{g}({\bf r}). \label{ggp2}
\end{eqnarray}
\end{mathletters}

As is customary, we now expand the deviation from the ground state in the 
difference
\begin{equation}
\delta \hat\Psi_i ({\bf r})=\hat\Psi_i ({\bf r})-\Psi_i^g ({\bf r})\qquad 
(i=1,2) \label{exp}
\end{equation}
which still preserves the operator character of the problem.

We can now proceed in two ways: One way substitutes the expansion (\ref{exp}) 
into the Hamiltonian (\ref{ham1}) and retains the terms to second order only. 
The linear terms in $\delta \hat\Psi_i$ are found to cancel because $\Psi_i^g$ 
solves the stationary equations (\ref{ggp1}) and (\ref{ggp2}). The eigenmodes 
of the remaining Bogoliubov problem are then obtained by diagonalizing the 
quadratic part to obtain the diagonal Hamiltonian in term of creation and 
annihilation operators $\{ a_\nu^\dagger,a_\nu\}$ and 
$\{ b_\nu^\dagger,b_\nu\}$, which reduce to the eigenmodes of the separated 
condensates when the mode coupling $v_3$ vanishes. In terms of these 
operators, the Hamiltonian takes the form of decoupled linear 
oscillator relations.

Alternatively, we may proceed from the equations of motion (\ref{tgp}) and 
linearize these. Finding the eigenmodes, then consists in 
looking for the eigenfrequencies of this linear time dependent problem. In 
the present paper we choose to approach the problem in this way.

From the two-component time dependent Gross-Pitaevskii equations (\ref{tgp})
we get the linearized equations
\begin{eqnarray}
i\hbar {\partial \over {\partial t}} \delta \Psi_{1}&=&[\hat H_{HO}^{(1)}+
2v_{1}\Psi_{1}^{g 2}]\delta \Psi_{1}
+v_{1}\Psi_{1}^{g 2}\delta\Psi_{1}^{*}
+v_{3}(\Psi_{2}^{g 2}\delta \Psi_{1}+
\Psi_{1}^{g}\Psi_{2}^{g}
\delta \Psi_{2}^{*}+\Psi_{1}^{g} \Psi_{2}^{g} \delta \Psi_{2} )\label{lin1} \\
i\hbar {\partial \over {\partial t}} \delta \Psi_{2}&=&[\hat H_{HO}^{(2)}+
2v_{2}\Psi_{2}^{g 2}]\delta \Psi_{2}+v_{2}\Psi_{2}^{g 2}\delta 
\Psi_{2}^{*}+v_{3}(\Psi_{1}^{g 2}\delta \Psi_{2}+
\Psi_{1}^{g}\Psi_{2}^{g}
\delta \Psi_{1}^{*}+\Psi_{1}^{g}\Psi_{2}^{g}
\delta \Psi_{1}) \label{lin2}
\end{eqnarray}
where 
\begin{eqnarray}
\hat H_{HO}^{(i)}=-{{\hbar^2}\over {2m}}\nabla^2+V_i ({\bf r})-\mu_i.
\end{eqnarray}
In writing these equations, we have utilized the fact that the ground state 
functions $\Psi_i^g$ have been chosen real. The time dependent linear 
equations (\ref{lin1}) and (\ref{lin2}) can be solved by generalizing the 
Bogoliubov transformation \cite{Fetter} to the two-condensate inhomogeneous 
case 
\begin{eqnarray}
\delta \Psi_1 ({\bf r})&=&\sum_\nu [\alpha_\nu^{(1)}({\bf r})
\hat a_\nu e^{-i E_\nu t/
\hbar}+\beta_\nu^{(1)}({\bf r})^* \hat a_\nu^\dagger  e^{i E_\nu t/
\hbar}] \label{bogo1} \\
\delta \Psi_2 ({\bf r})&=&\sum_\nu [\alpha_\nu^{(2)}({\bf r})
\hat b_\nu e^{-i E_\nu t/
\hbar}+\beta_\nu^{(2)}({\bf r})^* \hat b_\nu^\dagger  e^{i E_\nu t/
\hbar}], \label{bogo2}
\end{eqnarray}
where the sum over $\nu$ is a sum over elementary excitation modes of the 
system. Inserting this ansatz into (\ref{lin1}) and (\ref{lin2}) we find the 
double condensate Bogoliubov-de Gennes equations
\begin{mathletters}
\label{bdg}
\begin{eqnarray}
\left [ \hat H_{HO}^{(1)}+2v_{1}\Psi_1^{g 2}+
v_3\Psi_2^{g 2} \right ] \alpha_\nu^{(1)}+
v_3\Psi_1^g\Psi_2^g
(\alpha_\nu^{(2)}+\beta_\nu^{(2)})+
v_1 \Psi_1^{g 2} \beta_{\nu}^{(1)} &=& E_\nu \alpha_\nu^{(1)} \label{bdg1} \\
\left [ {\hat H_{HO}^{(1)}+2v_{1}\Psi_1^{g 2}+
v_3\Psi_2^{g 2}} \right ]\beta_\nu^{(1)}+v_3\Psi_1^g\Psi_2^g
(\alpha_\nu^{(2)}+\beta_\nu^{(2)})+
v_1{\Psi_1^{g 2}} \alpha_{\nu}^{(1)} &=& -E_\nu \beta_\nu^{(1)} \label{bdg2} \\
\left [ \hat H_{HO}^{(2)}+2v_{2}\Psi_2^{g 2}+
v_3\Psi_1^{g 2} \right ] \alpha_\nu^{(2)}+
v_3\Psi_1^g\Psi_2^g
(\alpha_\nu^{(1)}+\beta_\nu^{(1)})+
v_2{\Psi_2^{g 2}} \beta_{\nu}^{(2)} &=& E_\nu \alpha_\nu^{(2)} \label{bdg3} \\
\left [ \hat H_{HO}^{(2)}+2v_{2}\Psi_2^{g 2}+
v_3\Psi_1^{g 2} \right ] \beta_\nu^{(2)}+
v_3\Psi_1^g\Psi_2^g
(\alpha_\nu^{(1)}+\beta_\nu^{(1)})+
v_2{\Psi_2^{g 2}} \alpha_{\nu}^{(2)} &=& -E_\nu \beta_\nu^{(2)}. \label{bdg4} 
\end{eqnarray}
\end{mathletters}
The ensuing eigenvalue problem is nonhermitian, but it 
has a symmetry equivalent 
with that in the case of a single condensate: Taking the complex conjugate, 
exchanging the $\alpha$:s and $\beta$:s and reversing the sign of the 
eigenvalue $E_\nu$, we generate equations identical with (\ref{bdg}). 
This time reversal operation guarantees that the eigenvalues are real and come
 in positive and negative pairs. There is also one pair of eigenvalues which 
are zero. These eigenvalues correspond to the Gross-Pitaevskii solutions, and 
are not in this case interesting since they do not represent any oscillations. 
They correspond to the zero frequency collective modes discussed for
the single condensate by Lewenstein and You \cite{Lew}.

\section{Numerical methods}

\subsection{Solving the secular equations}

We now consider the case where we have spherically symmetric trapping 
potentials 
with $\Omega_{1}/\Omega_{2}=\sqrt{2}$. The coupled 
Eqs. (\ref{ggp}) and (\ref{bdg}) are then considerably simplified. 
In that case, the obvious and simplest way to solve the Bogoliubov-de 
Gennes equations (\ref{bdg}) is to discretize the solutions and the 
derivatives. 
This method works well if the problem can be reduced to one
dimension. Thus the radial part is solved for, which means 
that we 
can only see the so called breathing modes \cite{Jin96,Mew97} of the 
condensates. 
When discretizing the 
solutions $\alpha^{(i)},\beta^{(i)}$ and the derivatives we get an eigenvalue 
problem 
\begin{equation}
\left(
\begin{array}{c c c c}
{\hat M}_{1} & v_1 \Psi_1^{g 2} & v_3 \Psi_1^g \Psi_2^g 
&v_3 \Psi_1^g \Psi_2^g \\
- v_1 \Psi_1^{g 2}&-\hat M_1 &-v_3 \Psi_1^g \Psi_2^g &
-v_3 \Psi_1^g \Psi_2^g \\
v_3 \Psi_1^g \Psi_2^g &v_3 \Psi_1^g \Psi_2^g&\hat M_2 &
v_2 \Psi_2^{g 2}\\
-v_3 \Psi_1^g \Psi_2^g &-v_3 \Psi_1^g \Psi_2^g&- v_2 \Psi_2^{g 2}&
-\hat M_2
\end{array} 
\right)
\left(
\begin{array}{c} 
\alpha^{(1)} \\ \beta^{(1)} \\
\alpha^{(2)} \\
\beta^{(2)} 
\end{array}
\right)
=E 
\left(
\begin{array}{c}
\alpha^{(1)} \\
\beta^{(1)} \\
\alpha^{(2)} \\
\beta^{(2)} 
\end{array}
\right) \label{mat}
\end{equation}
with 
\begin{equation}
\hat M_i = \hat H_{HO}^{(i)}+2v_i \Psi_i^{g 2}+v_3  \Psi_j^{g 2} 
\qquad i\neq j.
\end{equation}
The eigenvalue problem then concerns a matrix of size $(4n)\times (4n)$, 
where $n$ is the number of gridpoints used to build up the eigensolutions.

The Bogoliubov-de Gennes equations can also be 
solved using a basis set method, where the solutions are expanded in some 
suitable orthonormal set of functions. In the case of harmonic traps, it is 
favourable to use the eigenstates of the harmonic oscillator. This gives an 
eigenvalue problem for the expansion coefficients instead of the more direct 
method of discretizing the solutions. Expanding 
the solutions in some basis set works particularly well when the 
condensates are separated by gravity. However, in the effective 
one-dimensional problem we consider, we found the grid method to be superior. 
In order to evaluate the matrix in Eq. (\ref{mat}) we need 
the solutions of the Gross-Pitaevskii equations (\ref{ggp}).
These were obtained with the method of steepest descent, which 
has been successfully used in earlier work \cite{Dalf,Esry2} 
on the nonlinear Schr\"odinger equation.

\subsection{Time dependent methods}

The numerical diagonalization of the 
secular equation can be checked by a simple time dependent method. 
Choosing an initial state which is not an eigenstate of Eq. (\ref{tggp3}) and 
(\ref{tggp4}) is expected to produce oscillations in the densities. 
Following the 
time evolution of these at an arbitrary point in the condensate and fourier 
transforming the corresponding signal gives us the spectrum of the 
eigenoscillations. This crude fourier method, where an initial state which 
slightly differs from 
the ground state is chosen, works in the spirit of kicking the condensates and 
letting them ring. It gives us the spectrum in one go but it is not very 
accurate since we have used only 50 gridpoints in the densities. It is not
obvious how to get a good spectrum with this method whithout greatly 
increasing the computing work. Increasing the grid to one hundred points 
has been 
found to give better agreement with the method in Sec. III.A, but the 
drawback is that the computing time 
becomes forty times longer. This method is thus found to be 
of limited accuracy 
and has to be considered as a check on the diagonalization procedure from Eq.
(\ref{mat}).

Another approach to calculating the eigenvalue spectrum is to follow the 
response
of the system to an external driving force \cite{KBG2}. This is most easily 
done directly on the nonlinear equations (\ref{tgp}) treated as 
coupled scalar equations. As discussed in Ref.\cite{KBG2}, we add to the 
external trapping potentials $V_i (r)$ the driving
\begin{equation}
V_d (r)=d \cos (kr+\omega_d t), \label{drive}
\end{equation}
where $\omega_d$ is a variable driving frequency, and $k$ is a suitably chosen 
scaling parameter, whose value does not greatly affect the results. We can 
then follow the response of the condensate densities as functions of the 
driving 
frequency $\omega_d$ by fourier transforming the 
corresponding changes in the densities as explained above. 
This method is more accurate and also
more transparent, since the condensate response is truly dramatic when 
$\omega_d$ approaches one of the resonance frequencies. One problem is
that with a sufficiently strong driving $d$, harmonic generation is
also seen at multiples of the true resonance frequencies. The
nonlinearities of the dynamic evolution mixes the frequencies.

\section{Results}

The results of our numerical calculations are reported in Fig. \ref{main}. 
The continuous lines are the modes obtained from the secular equation 
(\ref{mat}) as 
explained above. Because of the spherical symmetry, the equations are reduced 
to a one-dimensional eigenvalue problem, which means that we see only
the so called 
breathing modes of the condensates. At $v_3 =0$ the condensates do not 
interact with each other and behave like uncoupled systems. On the left 
hand side we number the eigenfrequencies as $\sharp 1$ to $\sharp 10$ 
starting from below. At $v_3=0$ the frequencies $\sharp 2,\sharp 4,\sharp 7$ 
and $\sharp 9$ belong to condensate one, the intertwined ones belong to 
condensate two. The spectrum is seen to display avoided crossings
between $\sharp 1$ and $\sharp 2$ or $\sharp 6$ and $\sharp 7$ around 
$v_3=0.07$.

The discrete points given in Fig.1 derive from the response of the kicked 
condensate as described in Sec. III.B. As explained there, the results are not 
very accurate but serve as an overall check on the spectrum.  
We found, that choosing a typical 
initial state to be the Gross-Pitaevskii solution with $v_3$ slightly shifted
 $(\Delta v_3=0.001)$ gives 
the best spectrum for levels up to $\sharp 4$. With this method it  
is possible to detect the four lowest frequencies without any major 
inaccuracy; for higher frequencies the agreement is less
satisfactory. The solution 
of the time dependent Gross-Pitaevskii equations have to be computed for long 
times in order to get a high resolution spectrum. This was done with the grid  
method as explained earlier. 
The results from the fourier method shown in Figs.\ref{spec1}-\ref{spec3} used
fifty gridpoints in the density. This gives stable iterations and allows us to
solve for long times. The frequencies of the spectrum are slightly higher than 
the spectrum 
calculated from Eqs. (\ref{mat}). Fig. \ref{hund} shows a portion of the 
spectrum in Fig. \ref{main} when we have increased the number of grid points 
to 100. We see
that the frequencies obtained uniformly tend to approach the results 
from the secular equations. The fortyfold increase in computer time, however, 
makes it costly to cover the full spectral range shown in Fig. \ref{main}. 

The eigenoscillations of the condensates are most 
clearly observed when the spectrum is investigated using the method of an 
external driving field. This method produces improved
results at least 
for the three lowest frequencies. In this paper we have used $d=0.01$ in 
order to 
keep $V_d<< v_i |\Psi_i^g|^2$ and set $k=1/2$; see Ref.\cite{KBG2}. 
In Fig. \ref{resp} we show the 
observed response of the two condensates at and near the lowest level with 
$v_3=0.07$. 
To the left, we see the dramatic increase of the magnitude of the response 
when $\omega_d$ is changed from $3.4$ to the resonance value $3.8$. As this 
calculation has been done in the region near the avoided crossing, 
we find the same 
frequency nearly equally strong 
in the fourier transforms of either condensate (to the right 
in Fig. \ref{resp}). In Fig. \ref{tva}, we show the radial densities
of condensate one and two. We see that, already for $v_3>0.03$, the
repulsive interaction tends to separate the condensates thus
decreasing their coupling.

In order to illuminate the behaviour of the coupled double
condensates, we consider the levels $\sharp 1$ and $\sharp 2$ near
$v_3 =0.07$, where an avoided crossing is suggested. Near $v_3 =0.0$,
level $\sharp 2$ is an excitation in condensate 1 and $\sharp 1$ is
condensate 2. If we look at the amplitudes $\alpha_1,\beta_1,\alpha_2$
and $\beta_2$ at $v_3=0.04$, Fig. \ref{ett}b shows that the dominating
oscillation amplitude is still $\alpha_2$; the amplitudes $\alpha_1$
and $\beta_1$ in Fig. \ref{ett}a are clearly smaller. If we compare
these amplitudes with the densities of the condensates,
Fig. \ref{tva}, we find that condensate 1 has no oscillational
amplitudes after $r\approx 2.5$, because the density goes to zero. The
condensate 2 reaches out to about $r\approx 3.5$ because it is repelled
by condensate 1. The features in the oscillation amplitudes $\alpha_1$
and $\beta_1$ around $r\approx 2$ are caused by the coupling to
condensate 2; for $\alpha_2,\beta_2$ the analogous effects are seen
near $r\approx 0$. Beacuse the coefficients in Fig. \ref{ett} are
eigenmodes of the coupled problem, there are areas where the
components nearly decouple; thus near $r\approx 2.2$, the oscillations
are nearly purely taking place in condensate 1.

When we follow the oscillational level $\sharp 1$ to $v_3 =0.08$,
Fig. \ref{tre}, we assume that an avoided crossing has been
passed. Thus the oscillational amplitudes have been transferred to
condensate 1, which is verified in Fig. \ref{tre}a. The amplitudes in
condensate 2, Fig. \ref{tre}b, are clearly smaller. They disappear
rapidly near $r\approx 0$, beacuse for this interaction strength,
condensate 2 is pushed well away from the center of the trap, see
Fig. \ref{tva}. Beacuse of the decreased overlap between the
condensates, they influence each others oscillation amplitudes far
less than for $v_3=0.04$.

Table I summarizes the results of all our numerical calculations. The 
frequencies of the five lowest eigenmodes are reported at $v_3=0.002$ and 
$v_3=0.02$. The first column is the result from the Bogoliubov-de Gennes 
secular equations (\ref{bdg}). The next column labelled 
$Fourier_1$ is obtained from the spectral response of the kicked condensates and the column $Fourier_2$ from the driven condensate response. Comparing the 
columns we obtain a good picture of the accuracy of our numerical methods. At 
$v_3=0.002$ the frequencies agree well. At $v_3=0.02$ the 
frequencies calculated with the Fourier methods are less satisfactory
as compared with 
the BdG-spectrum. Unfortunately, it is 
very easy to excite harmonic generations in the vicinity of level $\sharp 4$ 
and $\sharp 5$, which makes it difficult to find the eigenfrequencies 
for levels higher than $\sharp 4$. This may also explain the lack of
improvement in $Fourier_2$ over $Fourier_1$ at these frequencies.

\section{Conclusions}

We have generalized the Bogoliubov-de Gennes method to the case of two coupled
 condensates. The ensuing eigenvalues are solved for the spherically symmetric
excitations, the breathing modes. The calculations are numerically demanding, 
and obtaining the full spectrum in this way seems to be beyond the numerical 
capacity of our approach. We have also started from a symmetric ground state, 
even if we know that the condensates may separate due to symmetry breaking 
\cite{Gord}. This we have done mainly for numerical reasons, but our 
investigations of the condensate stability indicate that the range of 
interaction we are mainly interested in here, $v_3 \alt 0.03$, may lead to 
stable condensates located on top of each other. We report on these 
results in a separate communication \cite{me4}. 

We have used time dependent integrations to check the behaviour  of the 
energy eigenvalues in an indenpendent way. Starting from a perturbed state, 
the system evolution should contain all the eigenfrequencies of the system; 
here they are restricted to the symmetric ones of course. The advantage of 
this method is that a fourier transform brings out all frequencies at the 
same time. This method is found not to be very accurate. 
However, the method seems to verify the overall 
behaviour of the modes but in most regions the results 
fall above those based on the linearization method. This is, however, not true 
in all parameter ranges, see Fig. \ref{main}. 
A more exact, but quite time consuming 
method brings the results of the time integration closer to those derived 
from the linearized method (see Fig \ref{hund}).

By perturbing the condensates at a single frequency, the resulting response 
grows dramatically at the resonances (see Fig. \ref{resp}), but to obtain the 
full 
spectrum the calculations have to be repeated for all frequencies of interest. 
The fourier transform, however, allows one to localize the eigenfrequencies 
quite accurately (see Table I). 

Near the presumed level crossing, $v_3\approx 0.07$, the two
condensates are fully mixed, see Fig. \ref{spec3}. Here, however, the
lower frequency, $E\approx 3.8$, is much stronger than the upper one,
see Fig. \ref{resp}. This is clearly seen also at $v_3=0.04$ in
Fig. \ref{spec1}. The amplitudes of the kicked condensate, however,
depend on the point chosen for the fourier transform. For still
stronger couplings $v_3=0.08$, the two condensates are only weakly
coupled, see Fig. \ref{spec3}. This may be the result of decreasing
condensate overlap for large $v_3$; see Fig. \ref{tva}. The amplitudes
of the oscillations in the two condensates, as shown in
Fig. \ref{resp}, should not be directly compared; only the relative
strengths of the oscillational components are indicated. In general, we
find that level $\sharp 2$ is much harder to excite near $v_3 \approx
0.07$ than level $\sharp 1$ even if both are easily seen in our results.
The relative strengths of the two condensate
components in one oscillational mode varies with position as the
corresponding eigen amplitudes. For level $\sharp 1$, this is shown in
Figs. \ref{spec1},\ref{spec2} and \ref{spec3}. We point out that the
region where we have found strong mode mixing, $v_1\approx v_2 \approx
v_3$, does correspond approximately to the physical situation. This
suggests that our results may have experimental manifestations in real
systems.

\section{Acknowledgements}

We want to thank the Finnish Center for Scientific Computing for providing 
us with computer time.

\begin{figure}
\caption{The spectrum of the breathing modes. The lines are the 
eigenvalues from the Bogoliubov-de Gennes secular equations. 
The crosses (condensate 2) and the 
squares (condensate 1) is the spectrum calculated with the fourier
method. Throughout the 
calculations we have used $N_1=N_2=2100,v_1=0.02$ and $v_2=0.01$. All energies
are in units of $\hbar \sqrt{\Omega_1 \Omega_2}/2$. The levels are numbered 
at $v_3=0$ from $\sharp 1$ to $\sharp 10$ starting from below. The fourier 
method enables us to calculate the four lowest modes with a reasonable 
accuracy, since for higher frequencies the corresponding signal decreases 
rapidly.}
\label{main}
\end{figure}

\begin{figure}
\caption{As an example of our numerical data, this figure shows the 
spectrum computed by the fourier method at the levels 
$\sharp 1$ and $\sharp 2$. Here $v_3=0.04$ and the two condensates can
be seen to simultaneously oscillate with the frequency $E_1=3.1$
whereas level $\sharp 2$ is hardly seen at $E_2=4.6$. The relative
amplitude of the oscillations depends on the point where the density
is investigated; cf. Fig. \ref{ett}.}
\label{spec1}
\end{figure}

\begin{figure}
\caption{The same situation as in Fig.\ref{spec1} with $v_3=0.06$. 
The spectrum shows simultaneous oscillations of both condensates 
at level $\sharp 1$
with $E_1 =3.6$ and level $\sharp 2$ with $E_2 =4.8$.}  
\label{spec2}
\end{figure}

\begin{figure}
\caption{Same situation as in Fig.\ref{spec1} and Fig.\ref{spec2} 
with $v_3=0.08$ where the two condensates are decoupled for the two
lowest levels.}
\label{spec3}
\end{figure}

\begin{figure}
\caption{A comparison between the Bogoliubov-de Gennes spectrum and the 
fourier method spectrum calculated with a grid 
twice as dense as in the previous figures. 
The general trend with a denser grid is that the values approach those from 
the secular equation, at the 
expense of a fortyfold increase in computing time.}
\label{hund}
\end{figure}

\begin{figure}
\caption{The condensates respond dramatically when 
$\omega_d$ approaches a resonance frequency. The signals in the left part 
of the figure are the change in the density as function of time for two 
different 
driving frequencies. The fourier transform of these signals give the spectra
on the right. The time is
expressed in units of $2/\sqrt{\Omega_1 \Omega_2}$ and $E$ in 
$1/2 \hbar \sqrt{\Omega_1 \Omega_2}$. Since the interaction strength is here 
$v_3=0.07$ the responses of both condensates are
significant. Oscillations at frequency $\sharp 2$ are hardly seen here.}
\label{resp}
\end{figure}

\begin{figure}
\caption{The more weakly trapped condensate is forming a shell
  structure separating the condensates 
with increasing interaction strength $v_3$. Condensate one is
concentrated near the center of the trap $(r\approx 0)$ and condensate
two forms a shell around it.}
\label{tva}
\end{figure}

\begin{figure}
\caption{The amplitudes of the 
eigenmodes for level $\sharp 1$ from the Bogoliubov-de
  Gennes equations shown
  in a) for $\alpha^{(1)}$ and $\beta^{(1)}$, and in b) for 
$\alpha^{(2)}$ and $\beta^{(2)}$ at $v_3=0.04$.}
\label{ett}
\end{figure}

\begin{figure}
\caption{The same situation as in Fig. \ref{ett} with $v_3=0.08$. The
  roles of the two condensates are changed, which we interpret to
  derive from passing an avoided crossing.}
\label{tre}
\end{figure}

\begin{table}
\caption{A comparison between the Bogoliubov-de Gennes (BdG) solutions and the
fourier methods. Here 
Fourier$_1$ stands for the crude method whereas Fourier$_2$ is the method of
driving the condensate with a frequency $\omega_d$. The coalescing 
oscillations are not shown here.}
\begin{tabular}{c c c c c}
$v_3$ & BdG & Fourier$_1$ & Fourier$_2$ & Level $\sharp$ \\
\hline 
0.002 & 3.43& 3.55 & 3.50 & 1 \\
& 4.99 & 5.15 & 5.10& 2 \\
& 6.62 & 6.85 & 6.60 & 3 \\
& 9.34 & 9.65 & 9.60 & 4 \\
& 9.88 & 10.25 & 9.80 & 5 \\
\hline 
0.020 & 2.85 & 2.95 & 3.00 & 1 \\
& 4.59 & 4.80 & 4.70 & 2 \\
& 6.17 & 6.35 & 6.30 & 3 \\
& 8.54 & 8.90 & 9.00 & 4 \\
& 9.63 & 9.90 & 10.00 & 5 \\
\end{tabular}
\end{table}

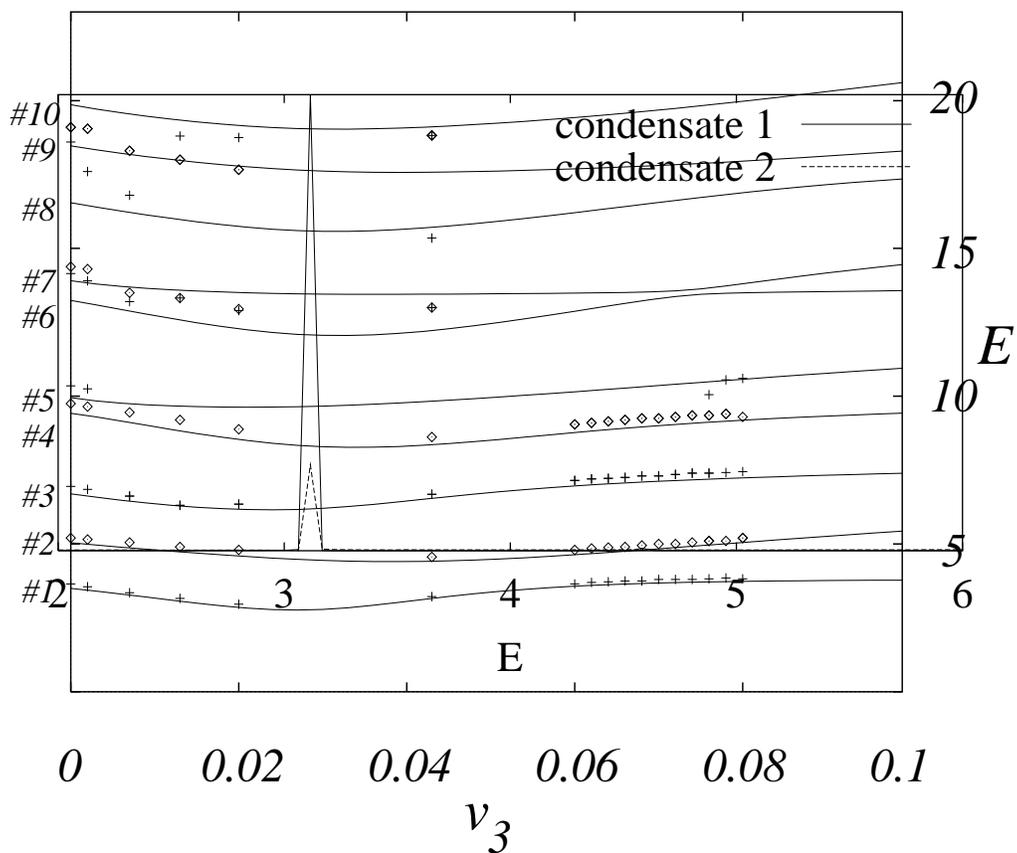
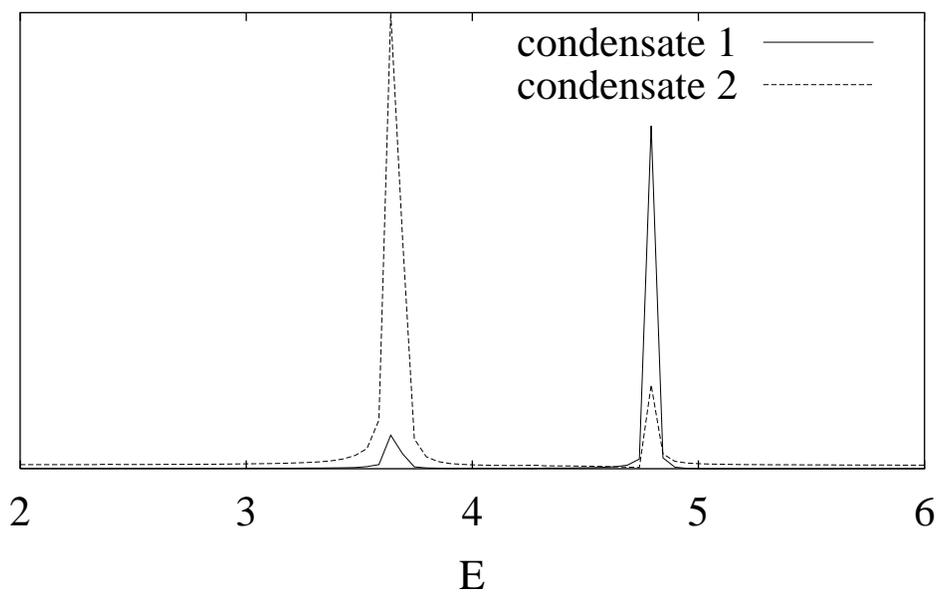
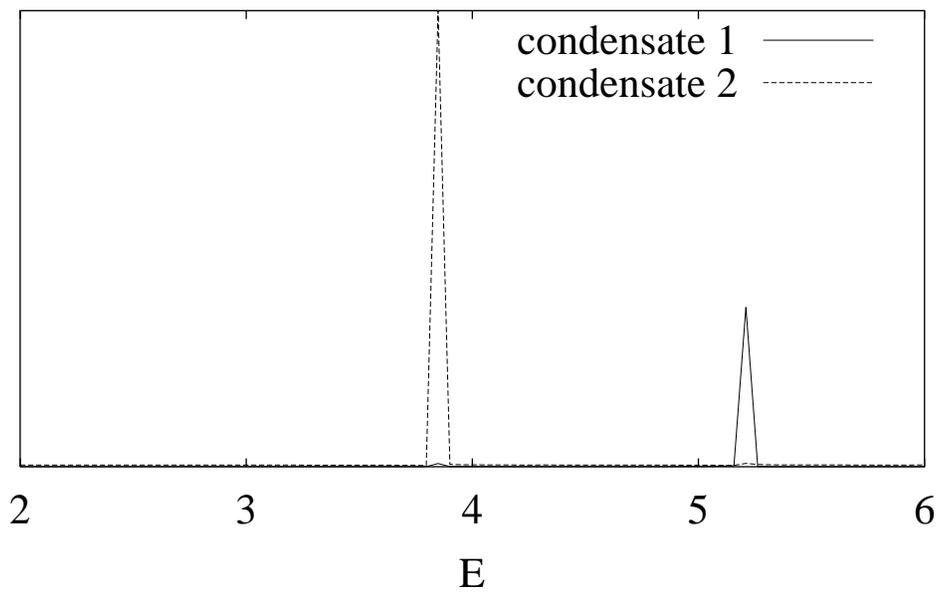
\end{document}